\documentclass{ifacconf}
\usepackage{graphicx}      
\usepackage{natbib}        
\usepackage{float}
\usepackage{subcaption}
\usepackage{amsmath}
\usepackage[boxruled, linesnumbered]{algorithm2e}
\usepackage{physics}
\usepackage{amssymb}
\usepackage{xcolor}
\usepackage{url} 
\let\oldnl\nl
\newcommand{\nonl}{\renewcommand{\nl}{\let\nl\oldnl}}
\newcommand{\rr}{\mathop{{\rm I}\mskip-4.0mu{\rm R}}\nolimits}
\newcommand{\Z}{\mathop{{\rm Z}\mskip-7.0mu{\rm Z}}\nolimits}
\newtheorem{remark}{\textbf{Remark}}
\newtheorem{problem}{\textbf{Problem}}
\newtheorem{definition}{\textbf{Definition}}

\begin{document}
\begin{frontmatter}
\title{An Observer-Based Key Agreement Scheme for Remotely Controlled Mobile Robots} 
\author{Amir Mohammad Naseri, Walter Lucia, Amr Youssef}
\address{Concordia Institute for Information Systems Engineering (CIISE)\\ Montreal, Quebec, Canada.\\
e-mail:\{amirmohammad.naseri,walter.lucia,amr.youssef\}@concordia.ca
}

\thanks[footnoteinfo]{This work is supported by the Natural Sciences and Engineering Research Council of Canada (NSERC).}

\begin{abstract}                
Remotely controlled mobile robots are important  examples of  Cyber–Physical Systems (CPSs). Recently, these robots are being deployed in many safety critical applications. Therefore,  ensuring their cyber-security is of  paramount importance. Different control schemes that have been proposed to secure such systems against sophisticated cyber-attacks require the exchange of secret messages between their smart actuators and the remote controller. Thus, these schemes require pre-shared secret keys, or an established Public Key Infrastructure (PKI) that allows for key agreement. Such cryptographic approaches might not always be suitable for the deployment environments of such remotely mobile robots. To address this problem, in this paper, we consider a control theoretic approach for  establishing a secret key between the remotely controlled robot and the networked controller without resorting to traditional cryptographic techniques. Our key agreement scheme  leverages a nonlinear unknown input observer and an error correction code mechanism to allow the robot to securely agree on a secret key with its remote controller. To validate the proposed scheme, we implement it using a Khepera-IV differential drive robot and evaluate its efficiency and the additional control cost acquired by it. Our experimental results confirm the effectiveness of the proposed key establishment scheme.
\end{abstract}

\end{frontmatter}
\section{Introduction}
Over the past recent years, the application of mobile robots in different safety critical domains, such as defence and space, search and rescue, health care, and industry 4.0, has gained an increasing interest \citep{tzafestas2013introduction, lewis2018autonomous, fragapane2020increasing}.  The research community has also been active in increasing the potential applications  of these robots in networked and distributed control systems setup \citep{santilli2021dynamic, klancar2017wheeled, liu2018formation, wang2021cloud}. 
On the other hand, such developments raise the concern of security and privacy of such systems \citep{dutta2021cybersecurity,li2022intelligent,Wang22TITS}. Mass adoption of robots leads to an increase in the possibilities of cyberattacks attacks against these systems. 

Different robotics cybersecurity issues, vulnerabilities threats, and risks have been  classified and discussed in
\citep{yaacoub2021robotics}. 
The target of an attacker can be any components of the Confidentiality, Integrity, Availability (CIA) triad on the level of the hardware of the robot, its firmware, or its communication channels \citep{yaacoub2021robotics}. Unlike cybersecurity of information technology systems, robots add the additional factor of physical interaction with the environment.
While taking the control of a desktop computer or a server may result in loss of information, taking the control of a robot may directly result in physical damages and endangering whatever or whoever is nearby. 
In this paper, we focus on applications where mobile robots are remotely controlled, i.e., where the  control inputs and sensor measurements are transmitted over insecure communication channels.

To guarantee any component of the CIA triad between the robot and the networked controller, or for the purpose of detection of different classes of cyberphysical attacks, in most of the proposed methods, sharing a secret key/seed is required.
For example, the authors of \citep{noura2018efficient} developed a physical-layer encryption algorithm for wireless machine-to-machine devices, in which sharing secret seeds is required for the implementation of the algorithm. On the other hand, the solution in \citep{noura2022novel} deals with the data integrity and source authentication problems, particularly for IoT devices. The proposed message authentication algorithm requires a secret seed/key to initialize the algorithm. 
Similarly, it is well-understood in the CPS community that to detect intelligent coordinated networked attacks such as covert attacks \citep{smith2015covert}, proactive detection actions must be coordinately taken in both sides of the communication channels \citep{ghaderi2020blended}. 
For example, moving-target \citep{griffioen2020moving} and sensor coding \citep{miao2016coding} based detection schemes implement such an idea to prevent the existence of undetectable attack, and both requires, for coordination purposes, that a secret seed is pre-shared between the plant and the controller.  
\color{black}
An anomaly detection scheme, specifically targeting differential-drive robots, is developed in \citep{cersullo2022detection}, where intelligent setpoint attacks are of interest. The proposed detector leverages two command governor modules and two pseudo-random number generators (each placed in one of the two sides of the network). It has been proved that such an architecture prevent the existence of undetectable setpoint attacks only if a shared seed between the two sides of the communication channel can be established.

%
From the above examples, it is clear that the key-establishment problem in cyber-physical systems, including mobile robots, is relevant for enhancing the security of such systems. 


\subsection{Background and Related Works}
%

Traditionally, key agreement is achieved through the use of symmetric or public key cryptographic protocols \citep{menezes2018handbook}. For example, 
using elliptic curve cryptography, in \citep{jain2021ecc}, the authors proposed a mutual authentication and key agreement scheme between
cloud-based robots (i.e., robots that access cloud resources)  and cloud servers.
However, such solutions might not always be usable for robotic systems.  Public key protocols are computationally demanding and require a public key infrastructure \citep{menezes2018handbook} and the support of a key revocation mechanism (e.g., see \citep{shi2021device}). These requirements make public key protocol impractical for robots with limited computational capabilities \citep{yaacoub2021robotics}.
On the other hand, symmetric key-based solutions assume the existence of a pre-shared key. However, the compromise of such long-term keys usually leads to compromising the security of the whole system.




%
%

Alternative key-establishing solutions leverage the seminal concept of wiretap channel introduced by Wyner in \citep{wyner1975wire}. Such schemes  are not based on traditional cryptographic mechanisms. Instead, they
utilize the role of noise, which is a natural characteristic in any communications system, to achieve secure communications. In particular, Wyner proved that if the communication channel between the sender and receiver is statistically better than the one from the sender to the eavesdropper, then it is possible to design an encoding mechanism to communicate with perfect secrecy. Over the years, such a concept has been leveraged to design different key-agreement protocols for CPSs, see, e.g., \citep{maurer1993secret, ahlswede1993common, lara2021key, sutrala2021authenticated,  zhang2017cross, rawat2017evaluating} and references therein. 
In \citep{maurer1993secret, ahlswede1993common}, a key-agreement protocol based on public discussion is proposed. 
In \citep{sutrala2021authenticated}, by considering a 5G-enabled industrial CPSs, a three-factor user authenticated key agreement protocol is developed; in \citep{zhang2017cross}, by using ambient wireless signals, a cross-layer key establishment model for wireless devices in CPSs is designed to allow devices to extract master keys at the physical layer. In \citep{rawat2017evaluating}, by exploiting an information-theoretic approach, the outage probability for secrecy rate in multipleinput multiple-output (MIMO) systems for CPSs is investigated.

While all the above solutions are developed for CPSs,  none of them takes advantage of the closed-loop dynamics of the underlying physical system dynamics to design the key agreement protocol. A first tentative to design a key agreement leveraging the physical properties of control systems can be found in \citep{li2011key}, where the authors exploited common information about the plant's state to establish a key between the sensor and the controller. However, the authors only consider the case where the eavesdropper cannot observe the plant's state.
%
%
%
%
More recently, in \citep{lucia2020wyner, lucia2021key}, control theoretical approaches have been proposed to design key-agreement scheme leveraging the asymmetry in the CPS model knowledge available to the defender and adversary. 

\subsection{Contribution}

Existing control-theoretical solutions targeting generic CPSs \citep{lucia2020wyner, lucia2021key} are developed under the assumptions of linear plant's dynamics (which is not the case for mobile robots) and they have never tested on a real testbed. Consequently, in a nutshell,  this work presents the following theoretical and practical contributions:
\begin{itemize}
    \item It extends the key-establishment solution in \citep{lucia2021key} to deal with the non-linear dynamics of mobile robots.
    \item It experimentally validates, using a remotely maneuvered Khepera IV\footnote{\url{http://www.k-team.com/khepera-iv}} mobile robot, the performance and the capacity of the proposed control theoretical key-agreement scheme.
\end{itemize}

\subsection{Notation and Paper Organization}

The set of real numbers and real-valued column vectors of dimension $n_r> 0$ are denoted with $\rr$ and $\rr^{n_r},$  respectively. $M\in \rr^{n_r \times n_c} $ denotes a real-valued matrix of size $n_r \times n_c.$ Moreover, $I_{n_1}\in \rr^{n_1\times n_1},$ $0_{n_0}\in \rr^{n_0\times n_0},$ and $1_{n_1}\in \rr^{n_1}$ denote the identity matrix, zero matrix, and all-ones column vector, respectively.
The sets of non-negative integer numbers and positive integer number 
is denoted by $\Z_+$ and $\Z_{>0}$. The transpose and inverse of matrix $M$ are denoted with $M^T$ and $M^{-1},$ respectively. Given a random variable $v \in \rr^{n_r}, v \sim \mathcal{N}(\mu_{n_r}, \Sigma_{n_r})$ indicates a random variable normally distributed with mean $\mu_{n_r} \in \rr^{m_r}$  and covariance matrix $\Sigma_{n_r} > 0 \in \rr^{n_r \times n_r}$. Given an event $E$, the probability of occurrence of such event is denoted with $P(E)$.  Given a binary string $s \in \{0, 1\}^{n_s},$ $s[i]$ denotes the $i-th$ bit of $s.$

The rest of the paper is organized as follows. In Section~\ref{sec:system_setup_problem _formulation}, first, the robot model  and the adversary are presented, then, the considered key-establishment problem is stated. In Section~\ref{sec:key_agreement_protocol}, the proposed protocol for the key agreement in described. Experimental results obtained using a Khepera IV differential-drive robot are presented in Section~\ref{sec:proof_of_concept_and_practical_results}. Finally, Section~\ref{sec:conclusion} concludes the paper with some final remarks.
\begin{figure}
    \centering
    \begin{subfigure}[b]{0.49\columnwidth}
        \centering
        \includegraphics[width=0.9\columnwidth]{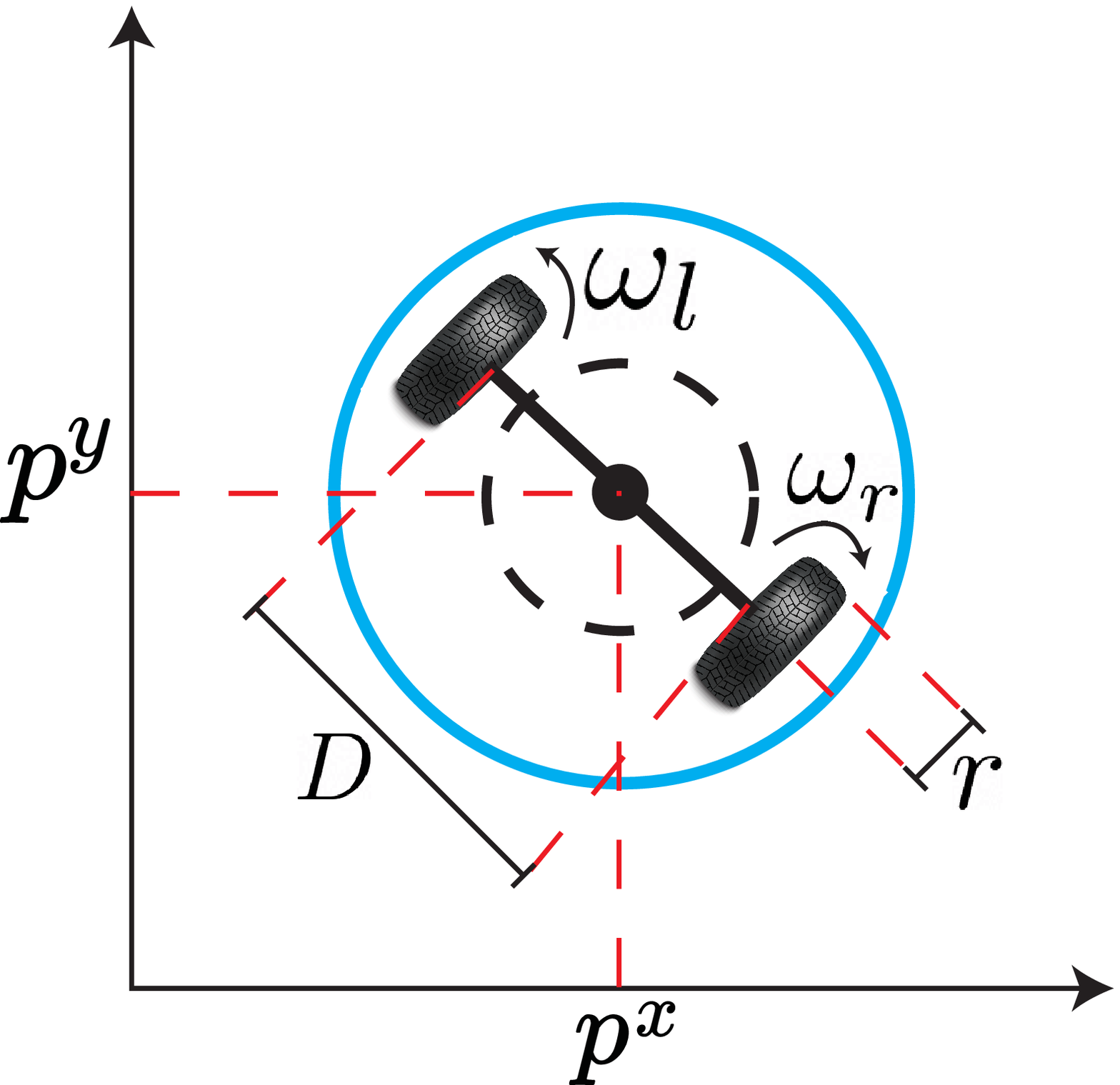}
        \caption{Differential Drive.}
        \label{fig:robot_differential_drive_mode}
    \end{subfigure}
    \hfill
    \begin{subfigure}[b]{0.49\columnwidth}
        \centering
        \includegraphics[width=0.9\columnwidth]{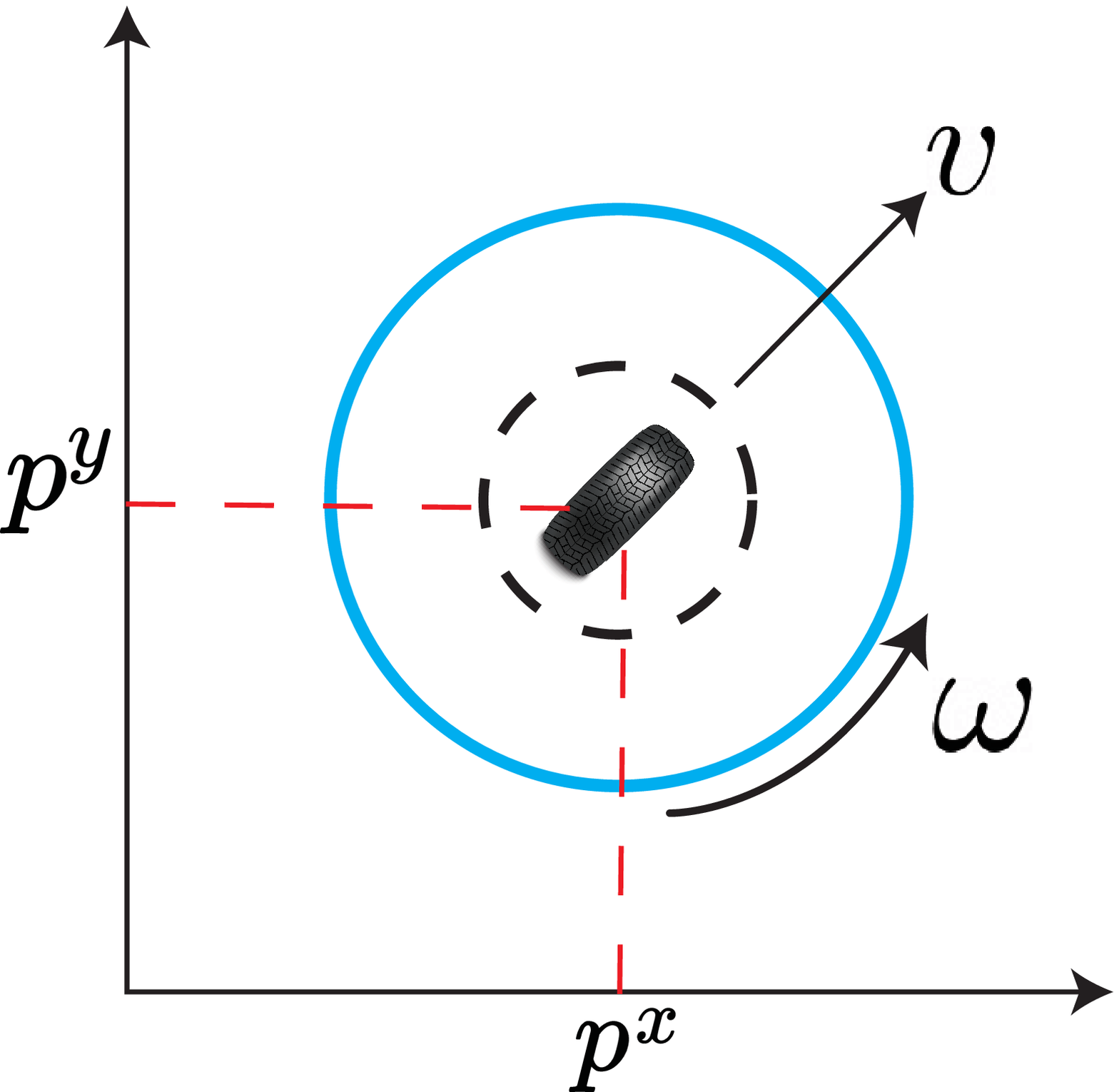}
        \caption{Unicycle.}
        \label{fig:robot_unicycle_mode}
    \end{subfigure}
    \hfill
    \vspace{-0.4cm}
    \caption{Differential-drive and unicycle models.}
    \label{fig:robot_schematic}
\end{figure}
\section{Preliminaries and Problem Formulation}
\label{sec:system_setup_problem _formulation}

\begin{definition}\label{def:ECC-linear}
Given three positive integers $n_c\in \Z_{>0}, k_c\in \Z_{>0}, d_c\in \Z_{>0},$ a linear Error Correcting Code (ECC) defines a linear transformation of a binary string $s \in \{0, 1\}^{k_c}$ into a subspace $\mathcal{C} \in \{0, 1\}^{n_c}$ of cardinality $2^{k_c}$ such that
\begin{itemize}
    \item $\forall (c_1, c_2) \in \mathcal{C}, c_1 \neq c_2$, the Hamming distance $d_H(c_1, c_2) < d_c.$
    \item the maximum number of errors that can be corrected is $\frac{d_c - 1}{2}.$
\end{itemize}
\hfill $\Box$
\end{definition}

In what follows, we consider a scenario where a mobile robot is manoeuvred by a networked controller and the network infrastructure is vulnerable to eavesdropping attacks.




\subsection{Robot Model}
\label{subsec. differential_drive_robot_model}

Among different existing categories of mobile robots, wheeled-mobile robots are very common for ground vehicles and they find application in different domains such as surveillance and warehouse automation.  Moreover, among the nonholonomic configurations, the differential-drive structure, characterized by two rear independently-driven wheels and one or more front castor wheels for body support, is often adopted in the industry \citep{martins2017velocity}. A schematic of a differential-drive robot is shown in Fig.~\ref{fig:robot_differential_drive_mode}. 

The pose of a differential-drive robot is described by the planar coordinates $(p^x,p^y)$ of its center of mass and orientation $\theta$ (see Fig.~\ref{fig:robot_differential_drive_mode}).
By resorting to the forward Euler discretization
method and a sampling time $T>0,$  the discrete-time kinematic model of the differential-drive is given by \citep{luca2001control}:
\begin{equation}\label{eq:robot_discrete_time_equation}
    \begin{array}{rl}
         p^x(k+1) = & p^x(k) \!+\! \frac{T r}{2}\cos{\theta(k)}(\omega_r(k) + \omega_l(k)) + \zeta^{p^x}(k)\vspace{0.1cm}\\
         p^y(k+1) = & p^y(k) \!+\! \frac{T r}{2}\sin{\theta(k)}(\omega_r(k) + \omega_l(k))+\zeta^{p^y}(k)\vspace{0.1cm}\\ 
         \theta(k+1) = & \theta(k) \!+\! \frac{T r}{D}(\omega_r(k) - \omega_l(k))+\zeta^{\theta}(k)\\
    \end{array}
\end{equation}
%
where $r>0$ is the radius of the wheels, $D>0$ the rear axle length, and $u^{D}=[\omega_r,$ $\omega_l]^T\in \rr^2$ the control input vector, which consists of the angular velocities of the right and left wheel, respectively. $\zeta(k)=[\zeta^{p^x}(k), \zeta^{p^y}(k), \zeta^{\theta}(k)]^T \sim \mathcal{N}(0, \mathcal{W})$ is the process noise with $\mathcal{W}\in \rr^{3\times 3}.$ 
Let $x(k)=[p^x(k), p^y(k), \theta(k)]^T\in \rr^3$ denote the robot's state vector. It is assumed that $x(k)$ can be estimated leveraging the measurement vector $y(k)\in \rr^{n_p},$ $n_p>0,$ obtained via odometric calculations and/or exteroceptive (e.g., sonar, laser) sensors \citep{d2015mobile}, i.e.,:
\begin{equation}\label{eq:robot_sensor_measurements}
    y(k) = h(x(k)) + \xi(k)
\end{equation}
where $h(x(k))$ denotes the nonlinear output equation, and  $\xi(k) \sim (0, \mathcal{V}),$ $\mathcal{V}\in \rr^{n_p \times n_p},$ the measurement noise, uncorrelated with $\zeta(k)$.

By denoting with $v(k)$ and $\omega(k)$ the linear and angular velocities of the center of mass of the robot, it is possible to apply to \eqref{eq:robot_discrete_time_equation} the transformation
\begin{equation}\label{eq:from_diff-drive_to_unicycle}
    \left[\begin{array}{c}
         v(k)  \\
         \omega(k) 
    \end{array}\right] = H \left[\begin{array}{c} \omega_r(k)\\
                         \omega_l(k)  \end{array}\right],\quad
                         H:=\left[\begin{array}{cc}
                            \frac{r}{2} &\frac{r}{2}\vspace{0.1cm}\\
                            \frac{r}{D} &\frac{-r}{D}
                         \end{array}\right] 
\end{equation}
and describe the robot behaviour by means of the following unicycle model (see Fig.~\ref{fig:robot_unicycle_mode}):
\begin{equation}\label{eq:robot_unicycle-model}
    \begin{array}{rl}
         {p}^x(k+1)=&{p}^x(k)+T v(k)\cos{\theta(k)}+ {\zeta}^{p^x}(k) \vspace{0.1cm} \\
         {p}^y(k+1)=&{p}^y(k)+T v(k)\sin{\theta(k)}+ {\zeta}^{p^y}(k) \vspace{0.1cm} \\
         {\theta}(k+1)=&{\theta}(k)+T\omega(k)+ {\zeta}^{\theta}(k)
    \end{array}
\end{equation}
where $u^{U}(k)=[v(k),\omega(k)]^T\in \rr^2$
is the control input vector of the unicycle.

\subsection{Adversary Model}
\label{subseq. Adversary_model}

We assume a passive adversary capable of eavesdropping the control input and sensor measurements transmitted between the plant and the networked controller, see Eve in Fig.~\ref{fig:protocol_system_setup}. We also assume that the adversary is aware that the robot is a differential-drive robot but it might not have exact knowledge of all the  robot's parameters (e.g.,  $T,r,D,\mathcal{W}$) and robot's measurement function (e.g., $h(\cdot)$ and $\mathcal{V}$). Therefore, we assume that the adversary has the following model:
\begin{equation}
\label{eq:eve_discrete_time_equation}
    \begin{array}{rl}
         p^x_a(k+1) = & \! p^x(k) \! +\! \frac{T_a r_a}{2}\cos{\theta_a(k)}(\omega_{r}(k) \!+\! \omega_{l}(k))\!+\!\zeta^{p^x}_{a}(k)\vspace{0.1cm}\\
         p^y_a(k+1) = & \! p^y(k) \!+\! \frac{T_a r_a}{2}\sin{\theta_a(k)}(\omega_{r}(k) \!+\! \omega_{l}(k))\!+\!\zeta^{p^y}_{a}(k)\vspace{0.1cm}\\ 
         \theta_a(k+1) = & \! \theta_a(k) \!+\! \frac{T_a r_a}{D_a}(\omega_{r}(k) \!-\! \omega_{l}(k))+\zeta^{\theta_a}_{a}(k)\vspace{0.1cm}\\
         y_a(k) = & h_a(x_a(k))\!+\!\xi_a(k)
    \end{array}
\end{equation}
where $\zeta_{a}=[\zeta^{p^x}_{a}(k),$  $\zeta^{p^y}_{a}(k), \zeta^{\theta}_{a}(k)]^T \sim (0, \mathcal{W}_a),$  $\xi^{x}_{a}(k) \sim (0, \mathcal{V}_a),$ and $(T_a, r_a, d_a, h_a(\cdot), \mathcal{W}_a, \mathcal{V}_a)$ are the adversary estimations for the robot's model \eqref{eq:robot_discrete_time_equation}-\eqref{eq:robot_sensor_measurements}.

\begin{assum}
\label{assumption:eve_non_perfect_model}
Let $\mathcal{M} = \{T,r, D, \mathcal{W}, h(\cdot), \mathcal{V}\}$ and $\mathcal{M}_a=\{T_a,r_a, D_a,  \mathcal{W}_a, h_a(\cdot), \mathcal{V}_a\}$ be the robot's model knowledge available to the controller's designer and to the adversary, respectively. Then, 
\begin{equation}
    \label{eq:eve_non_perfect_model}
    \mathcal{M}  \neq \mathcal{M}_a
\end{equation}
\end{assum}

\begin{remark}
The model discrepancy \eqref{eq:eve_non_perfect_model} might arise for different reasons. First, the adversary might not be aware of the robot construction parameters $r,D$ or the output function $h(\cdot).$ Instead, the attacker might just be able to estimate them using identification techniques or by inspection (e.g., via cameras). Second,  while the defender can estimate $\mathcal{W},\mathcal{V}$ by performing offline experiments, see, e.g., \citep{antonelli2007linear,d2015mobile},  the eavesdropper can only perform online identification procedure relying on the online robot operations, which might be unsuitable for system identification purposes. 
\end{remark}

\subsection{Problem Formulation}
\label{subsec. Problem_formulation}

The here considered key-agreement problem can be stated as follows.

\begin{problem}
Consider the robot and adversary models \eqref{eq:robot_discrete_time_equation}-\eqref{eq:eve_non_perfect_model}. Without resorting to traditional cryptographic schemes, design a key agreement protocol between the robot and the networked controller such that the keys of length $n>0$ identified by the controller ($\mathcal{K}_c \in \{0, 1\}^n$), robot ($\mathcal{K}_r \in \{0, 1\}^n$) and attacker ($\mathcal{K}_a \in \{0, 1\}^n$) are such that
\begin{equation}
  P\{\mathcal{K}_c\ = \mathcal{K}_r \} \approx 1 \text{ and } P\{\mathcal{K}_c\ \neq \mathcal{K}_a \} \approx 1  
\end{equation}
\end{problem}

\section{Key Agreement Protocol}
\label{sec:key_agreement_protocol}

\begin{figure}[h!]
	\centering
	\includegraphics[width=0.9\columnwidth]{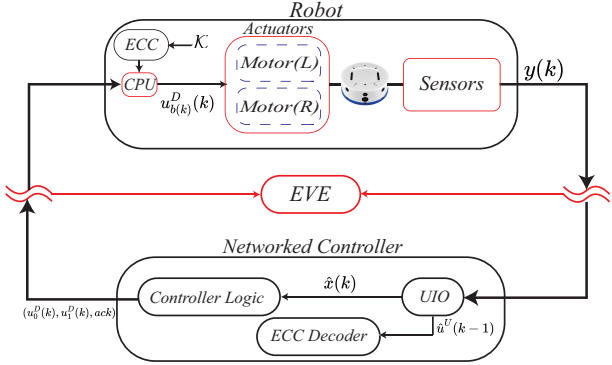}
	\caption{Control architecture for the proposed key agreement protocol.}
	\label{fig:protocol_system_setup}
\end{figure}

As proved in \citep{lucia2020wyner}, the asymmetry \eqref{eq:eve_non_perfect_model}  in the plant model knowledge  is sufficient to ensure the existence of a Wyner wiretap-like channel in networked cyber-physical systems. The latter is here leveraged to design an encoding mechanism for the considered key-exchange problem. In particular, the proposed key-agreement protocol is developed under the following assumptions.

\begin{assum}\label{assum:NUIO_of_the_robot}
The available sensor measurements are sufficiently rich to allow the existence of an Unknown Input Observer (UIO) capable of simultaneously estimate $x(k)$ and $u^D(k)$ from the available measurement vector $y(k).$  By denoting with $\hat{x}(k)$ and $\hat{u}^D(k)$ the estimated vectors, the UIO is abstractly modeled as the following recursive function
\begin{equation}\label{eq:NUIO_abstract_form}
    [\hat{u}^D({k-1}), \hat{x}(k)] =  UIO(u^D(k), \hat{x}({k-1}), y(k), \mathcal{M})
\end{equation}
where $(\hat{u}^D({k-1}), \hat{x}(k))$ and $(\hat{u}^D({k-2}), \hat{x}({k-1}))$ are the available estimations at time steps $k$ and $k-1$, respectively. Moreover, the eavesdropper is able to run the same UIO as in \eqref{eq:NUIO_abstract_form} with $\mathcal{M}_a$ instead of $\mathcal{M}.$
\end{assum}

In the sequel, we assume that the robot is equipped with a tracking controller which provides the control vector $u^U(k),\forall\,k,$ i.e.,
\begin{equation}\label{eq:generic_tracking_controller}
    u^U(k)=
    \left[ 
    \begin{array}{c}
         v(k)  \\
         \omega(k) 
    \end{array}
    \right]
    =
    f_c(x(k),x_r(k),\dot{x}_r(k),\ddot{x}_r(k))
\end{equation}
where $f_c(\cdot,\cdot,\cdot)$ denotes a generic controller, $x_r(k)\in \rr^3,\dot{x}_r(k)\in \rr^3,\ddot{x}_r(k)\in \rr^3$ are the reference state, velocity and acceleration vectors, respectively \citep{luca2001control}.

By referring to the networked control system architecture illustrated in Fig.~\ref{fig:protocol_system_setup}, the idea behind the proposed key-agreement protocol can be described in four points:

\textit{(P1)} -  The controller computes $u^U(k)$ as in  \eqref{eq:generic_tracking_controller}. Then, it generates two perturbed control inputs, namely $u_0^U(k)\in \rr^2$ and $u_1^U(k)\in \rr^2,$ by adding and subtracting a small bias vector $\Delta \in \rr^2$ to $u^U(k),$ i.e.,
\begin{equation}\label{eq:u1_u2_unicycle}
    u_0^U{(k)} = u^U(k)+ \Delta, \quad   u_1^U{(k)} = u^U(k) -\Delta 
\end{equation}
where $\Delta=[\Delta_v,   \Delta_{\omega}]^T,$  $\Delta_v\geq 0,$ $\Delta_{\omega} \geq 0$ and such that $\Delta_v+\Delta_{\omega}>0$ (i.e., at least one between $\Delta_v$ and $\Delta_{\omega}$ must be strictly greater than zero). Finally, the differential-drive control inputs are computed as $u_0^D(k)=H^{-1}u_0^U(k)$ and $u_1^D(k)=H^{-1}u_1^U(k)$, see  \eqref{eq:u1_u2_unicycle}, and the pair $(u_0^D(k),u_1^D(k))$ is sent to the robot.

\textit{(P2)} - Once the robot receives  $(u_0^D(k),u_1^D(k)),$ its CPU unit is in charge of deciding which one of the two control inputs should be used. To this end, it generates a random bit $b(k) \in \{0, 1\}$ and send to the actuators $u_{b(k)}^D(k).$ Note that the bit $b(k)$ and, consequently, the control signal applied to the robot ($u_{b(k)}^D(k)$) are unknown to the networked controller and to the eavesdropper. At each iteration, the robot appends $b(k)$ to the local key $\mathcal{K}_r.$

\textit{(P3)} - When the networked controller receives $y(k),$ it can run the UIO \eqref{eq:NUIO_abstract_form} and obtain the estimated pair $(\hat{x}{(k)}, \hat{u}^D{(k-1)}).$ Moreover, since also the pair $(u^D_0(k-1),u^D_1(k-1))$ is known, the controller can estimate the random bit $b(k-1)$ (used by the robot) as 
  \begin{equation}
        \label{eq:controller_bit_estimation}
        \hat{b}{(k-1)} = \begin{cases} 0 & \textit{if  } d_0 < d_1 \\ 1 & \textit{if  } d_1 < d_0 \end{cases}
    \end{equation}
where $d_0$ and $d_1$ are the distances between the estimated control input $\hat{u}^D{(k-1)}$ and   $(u_0^D{(k-1)},u_1^D{(k-1)}),$ i.e.,
\begin{equation}\label{eq:distance_estimation}
    \begin{array}{rl}
        d_0(k-1) &= \|\hat{u}^D{(k-1)} - u_0^D{(k-1)} \|_2,\vspace{0.1cm}\\
        d_1(k-1) &= \|\hat{u}^D{(k-1)} - u_1^D{(k-1)} \|_2
    \end{array}
\end{equation} 
At each iteration, the networked controller appends $\hat{b}(k)$ to the local key $\mathcal{K}_c.$

\textit{(P4)} - The adversary can run the  
UIO \eqref{eq:NUIO_abstract_form} with $\mathcal{M}_a$ instead of $\mathcal{M}$ and obtain a local estimation, namely $\hat{b}_a(k-1),$  of $b(k-1),$ to append to its local key $\mathcal{K}_a.$ However, given the model discrepancy \eqref{eq:eve_non_perfect_model}, the covariance of the unknown input estimation error for the attacker is expected to be larger of the one obtained by the networked controller \citep{lucia2021key}. Consequently,
for a proper choice of $\Delta,$ it is expected that $P\{\mathcal{K}_c\ = \mathcal{K}_r \} \approx 1 \text{ and } P\{\mathcal{K}_c\ \neq \mathcal{K}_a \} \approx 1.$

Note that the above described UIO-based decoding scheme might not be robust against possible model mismatches and/or process and measurement noises. To make the protocol more robust, we enhance its decoding operations by means of an Error Correcting Code (ECC) scheme and a feedback acknowledgment signal, namely $ack,$ which is sent by the controller along with the pair of control inputs. 

By assuming, for the sake of simplicity and clarity, a linear ECC, the ECC and $ack$ feedback signal are used as follows (refer to Definition~\ref{def:ECC-linear} for the used notation and terminology):
\begin{itemize}
    \item The robot splits a randomly generated local key $\mathcal{K}$ into a sequence of substring $s_i.$ Each $s_i$ is
    encoded into a sequence of codewords $c_i$. Each bit of $c_i,$ namely $c_i[j],$ is sequentially used to decide  $b(k)$ in \textit{(P2)},  i.e., $b(k)=c_i[j].$     
    \item The robot estimates  $\hat{b}(k)$ as in \textit{(P3)} and collects them to obtain an estimation of the codewords $c_i,$ namely $\hat{c}_i.$ Then, the Hamming distance $d_{\hat{c}_i}$ is evaluated
    \begin{equation}\label{eq:HammingDistance}
        d_{\hat{c}_i}=\arg\min_{c\in \mathcal{C}}d_H(c,c_i)
    \end{equation}
    %
    If $d_{\hat{c}_i}$ is much smaller than the number of correctable errors, then the codeword is accepted, the binary string $\hat{s}_i$ (associated to $\hat{c}_i$ via ECC) is appended $\mathcal{K}_c$, and a positive $ack_i=1$ is sent. Otherwise, the codeword is discarded and  $ack_i=0$ is sent.
    \item The robot, for every received $ack_i=1,$ append  ${c}_i$ to $\mathcal{K}_r.$ 
\end{itemize}

The complete key-agreement protocol is summarized in Algorithm~\ref{alg:proposed_protocol}.
\SetKwComment{Comment}{/*}{ */}
\setlength{\algomargin}{1.5em}
{\LinesNumberedHidden
\begin{algorithm}[h!]
\DontPrintSemicolon
    \Comment*[l]{Robot}
    
    \ShowLn Initialization: {Generate $\mathcal{K},$ and set $\mathcal{K}_r=\emptyset.$}
    \ShowLn Split $\mathcal{K}$ into sub-strings $s_i\in \{0, 1\}^{k_c}$ \;
    \ShowLn Sequentially encode each $s_i$ is into codewords $c_i\! \in\! \{0, 1\}^{n_c}\!\in \mathcal{C}$\;
    \ShowLn At each time step $k$:\\
    - Sequentially use each bit $c_i[j]$ of $c_i$  to pick $b(k)=c_i[j]$ and
    apply to the robot $u^D_{b(k)}(k)$\; 
    %
    - When all $n_c$ bits of $s_i$ are used, the robot receives   $ack\in\{0,1\}$ from the controller\; 
     \eIf{$ack==1$}
     {
       $s_i$ is appended to $\mathcal{K}_r$\;
     }{
         $s_i$ is discarded \;
     }
   
    \BlankLine
    \BlankLine
    
    \Comment*[l]{Controller}
    \setcounter{AlgoLine}{0}
    \ShowLn Initialization: {Set $\mathcal{K}_c=\emptyset.$}\\
    \ShowLn At each time step $k:$ \\ 
    - the pair $(\hat{u}^D({k-1}), \hat{x}(k))$ and $\hat{b}{(k-1)}$ are estimated using \eqref{eq:NUIO_abstract_form} and \eqref{eq:controller_bit_estimation}, respectively.\\
    - $\hat{b}{(k-1)}$ is appended to the estimated codeword $\hat{c}_i$\;
    - When $n_c$ bits of $\hat{c}_i$ are estimated, the distance $d_{\hat{c}_i}$ is computed using \eqref{eq:HammingDistance}. \;
    %
    \eIf{$d_{\hat{c}_i} \ll \frac{d_c-1}{2}$}
    {
        The codeword $\hat{c}_i$ is considered valid\;
        $\hat{s}_i$ is decoded from $\hat{c}_i$ and appended to $\mathcal{K}_c$; send $ack=1$ \;
    }{
        \setcounter{AlgoLine}{0}
        The codeword $\hat{c}_i$ is considered invalid and discarded; send $ack=0$ \;
    }
    - Compute $(u_0^D(k),u_1^D(k))$ as in \eqref{eq:u1_u2_unicycle} and  send it 
\caption{Proposed Key Agreement Protocol }
\label{alg:proposed_protocol}
\end{algorithm}
}

\begin{remark}\label{remark:key-amplification}
The bias $\Delta$ in \eqref{eq:u1_u2_unicycle} and the ECC parameters $(n_c,k_c,d_c)$ are design parameters that can be tuned to achieve $P\{\mathcal{K}_c\ = \mathcal{K}_r \} \approx 1.$  Moreover, to ensure the correctness of the exchanged key, the controller and the robot can always publicity verify
its correctness exchanging the hash values associated to $\mathcal{K}_c$ and $\mathcal{K}_r.$
Moreover,
to eliminate the partial key knowledge gained by the adversary, the controller and the robot can also enhance the security of the exchanged key by means of standard privacy amplification procedures, see, e.g., \citep{van2006quantum,bennett1995generalized}. 
\end{remark}

\section{Experimental Results}
\label{sec:proof_of_concept_and_practical_results}

\begin{figure}[h!]
	\centering
	\includegraphics[width=0.85\columnwidth]{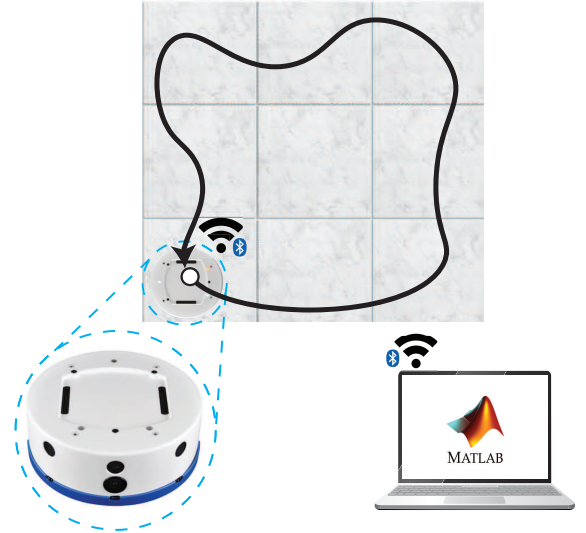}
	\caption{Experimental setup.}
	\label{fig:Experimental_utilized_setup}
\end{figure}

In this section, the effectiveness of the proposed key-agreement protocol  is verified by means of the experimental setup shown in Fig.~\ref{fig:Experimental_utilized_setup}. The setup consists of:
\begin{itemize}
    \item A laptop where a tracking controller is implemented in Matlab.
    \item A Khepera IV differential-drive robot.
    \item A Bluetooth 4.0 communication channel between the robot and the laptop for the two-way exchange of data, i.e., control inputs and sensor measurements.
\end{itemize}


\subsection{Khepera IV robot}

The Khepera-IV robot, produced by K-Team, is a differential drive robot whose discrete-time kinematic model is as in \eqref{eq:robot_discrete_time_equation},
where, $r=0.021\, [m]$ and $D = 0.1047\, [m],$ $\mathcal{W}=10^{-2}I_3,$ and the maximum angular velocities of the wheels is of $38\,[rad/\sec].$ On the other hand, the used measurement vector $y(k)$ consists of the wheels encoder measurements that, via odometric calculations, allow to obtain an estimation of the entire state of the robot \citep{luca2001control}. Consequently, the output equation \eqref{eq:robot_sensor_measurements} is modeled as $y(k)=x(k)+\xi(k)$ with $\xi(k)$ a Guassian noise with covariance matrix  $\mathcal{V}=10^{-4}I_3.$

In the performed experiments, the robot's processing unit is equipped with a server that receives and sends, via Bluetooth, the control inputs and sensor measurements. The used sampling time is $T_s = 0.2\,[\sec].$


\subsection{UIO, tracking Controller, and reference trajectoy}\label{sec:uio_controller_reference}

\textit{UIO:} The unicycle model \eqref{eq:robot_unicycle-model} under the control law \eqref{eq:u1_u2_unicycle} can be re-written (for compactness) as  
\begin{equation}\label{eq:model_for_UIO}
 x(k+1)=f(x(k),u^U(k)+\Delta_u)+\zeta(k),   
\end{equation}
with $\Delta_u$ the unknown bias of value $\pm \Delta.$ Then, the extended Kalman filter with unknown input estimation algorithm proposed in \citep[Appendix A]{guo2018detection} has been used to implement the UIO module~\eqref{eq:NUIO_abstract_form}. For completeness, the UIO operations, adapted to the considered setup, are reported in Algorithm~\ref{alg:NUIO}, where, $P_0^x=0_{3\cross3}$, $\hat{x}_0=[0, 0, 0]^T$, and $A_k,$ $B_k,$ $G_k$ are the matrices characterizing the linearization $x(k+1)=A_k x(k)+B_ku^U(k)+G_k\Delta_u(k)$ of \eqref{eq:model_for_UIO} along the state and input trajectories, i.e.,
\begin{equation}
    \label{eq:NUIO_required_linearization}
    \begin{array}{c}
         A_{k} \triangleq \eval{\pdv{f}{x}}_{(\hat{x}_{k|k}, u^{U}{(k)})},\,\,\, 
         B_{k} \triangleq \eval{\pdv{f}{u}}_{(\hat{x}_{k|k}, u^U{(k)})},\vspace{0.1cm}\\
         G_{k} \triangleq \eval{\pdv{f}{\Delta_u}}_{(\hat{x}_{k|k}, u^D{(k)})}
    \end{array}
\end{equation}
Consequently, 
$\hat{u}^{D}(k-1)=H^{-1}(u^U(k)+\hat{\Delta}_u(k-1)).$

\setlength{\algomargin}{1.5em}
\begin{algorithm}
\DontPrintSemicolon
    \KwIn{$u({k-1}), \hat{x}_{k-1}, y(k)$}
    \KwOut{$\hat{x}_{k}, \Delta_u({k-1})$}
    \BlankLine
    \Comment*[l]{Input Estimation}
    $\tilde{P}_{k-1} = A_{k-1}P_{k-1}^x(A_{k-1})^T+\mathcal{W}$\;
    $\tilde{R}_{k}^* = \tilde{P}_{k-1}+\mathcal{V}$\;
    $\Xi_{k} = (G_{k-1})^T(\tilde{R}_{k}^*)^{-1}$\;
    $M_{k}=(\Xi_{k}G_{k-1})^{-1}\Xi_{k}$\;
    $\hat{\Delta}_u({k-1}) = M_{k}(y({k}) - f(\hat{x}_{k-1}, u({k-1})))$\;
    $P_{k-1}^a = M_{k}\tilde{R}_{k}^*(M_{k})^T$\;
    \BlankLine
    \Comment*[l]{State Prediction}
    $\hat{x}_{k|k-1} = f(\hat{x}_{k-1}, u({k-1}) + \hat{\Delta}_u({k-1}))$\;
    $\Phi_{k} = (I - G_{k-1}M_{k})$\;
    $\bar{A}_{k-1} = \Phi_{k} A_{k-1}$\;
    $\bar{Q}_{k-1} = \Phi_{k}Q_{k-1} (\Phi_{k})^T + G_{k-1}M_{k}R_{k} (M_{k})^T (G_{k-1})^T$\;
    $P_{k|k-1}^x = \bar{A}_{k-1} P_{k-1}^x (\bar{A}_{k-1})^T + \bar{Q}_{k-1}$\;
    \BlankLine
    \Comment*[l]{State Estimation}
    $\Gamma_{k} = G_{k-1}M_{k}$\;
    $\tilde{R}_{k} = P_{k|k-1}^x + R_{k} + \Gamma_{k}R_{k} + R_{k}(\Gamma_{k})^T$\;
    $L_k = P_{k|k-1}^x+R_{k}(M_{k})^T(G_{k-1})^T)^T\tilde{R}_{k}^{-1}$\;
    $\hat{x}_{k} = \hat{x}_{k|k-1} + L_k(y(k) - h_2(\hat{x}_{k|k-1}))$\;
    $\Psi_k = I - L_k$\;
    $P_k^x = \Psi_k P_{k|k-1}^x \Psi_k^T + L_kR_k(L_k)^T - \Psi_k G_{k-1}M_kR_k(L_k)^T - L_kR_k(M_k)^T(G_{k-1})^T(\Psi_k)^T$\;
\caption{\footnotesize Non-Linear Unknown Input Observer }
\label{alg:NUIO}
\end{algorithm}

\textit{Tracking controller:} The robot's is controlled using the nonlinear controller based on  dynamic feedback linearization  described in \citep[Eq. 5.18]{luca2001control}.  
%
%
In the performed experiments, the controller has been implemented in Matlab using $k_{p}^x=k_{p}^y=1.10$, and $k_{d}^x=k_{d}^y=0.80.$

\textit{Reference Trajectory:}
The reference signal is the square-shaped trajectory shown in Fig.~\ref{fig:robot_trajectory_experiment}. 
The square's vertices are $\{(0,0), (1,0), (1,1), (0,1)\}$
and the timing laws for $(p_r^x,p_r^y),$ $(\dot{p}^x_{r}, \dot{p}^y_{r}),$ $(\ddot{p}^x_{r},\ddot{p}^y_{r})$ have been obtained using the built-in Matlab function  \textit{cubicpolytraj} which has been configured to travel each side of the square in $17\,[\sec].$ In the performed experiments, the square trajectory repeats three consecutive times. 

%

\subsection{Perturbed control inputs and ECC configuration}

\textit{Perturbed control inputs:} The pair $(u_0^U(k),u_0^U(k))$ has been obtained adding a small perturbation only into the linear velocity command $v{(k)}$ computed as in  \eqref{eq:khebera_control_law_deluca}, i.e., $\Delta_v>0$ and $\Delta_{\omega}=0,$ see \eqref{eq:u1_u2_unicycle}.

\textit{ECC configuration:} A simple repetition code has been used to implement the ECC. Therefore, the string $s_i$ consists of a single bit of $\mathcal{K}$ (i.e., $k_c=1$) and the codewords $c_i$ are vectors repeating $s_i$ for $n_c$ times. In the performed experiments, we set $n_c=3$ and a codeword is accepted only if the number of decoding errors $d_{\hat{c}_i}=0.$

\subsection{Results}
\begin{figure}[h!]
	\centering
	\includegraphics[width=0.95\columnwidth]{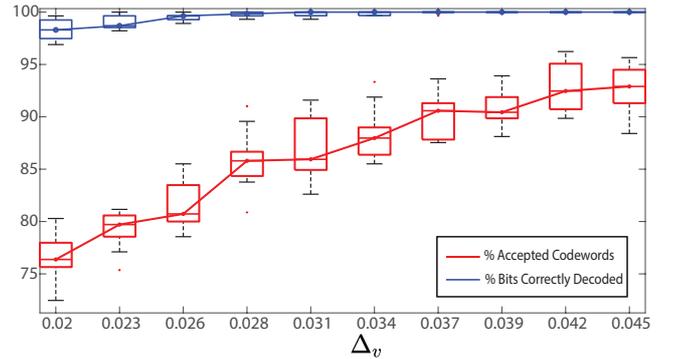}
	\caption{Percentage of accepted codewords (red boxplot) and correctly decoded bits (blue boxplot) by the controller for $\Delta_v \in [0.02, 0.045].$}
	\label{fig:Blocks_accepted_correct}
\end{figure}

\begin{figure}[h!]
	\centering
	\includegraphics[width=0.95\columnwidth]{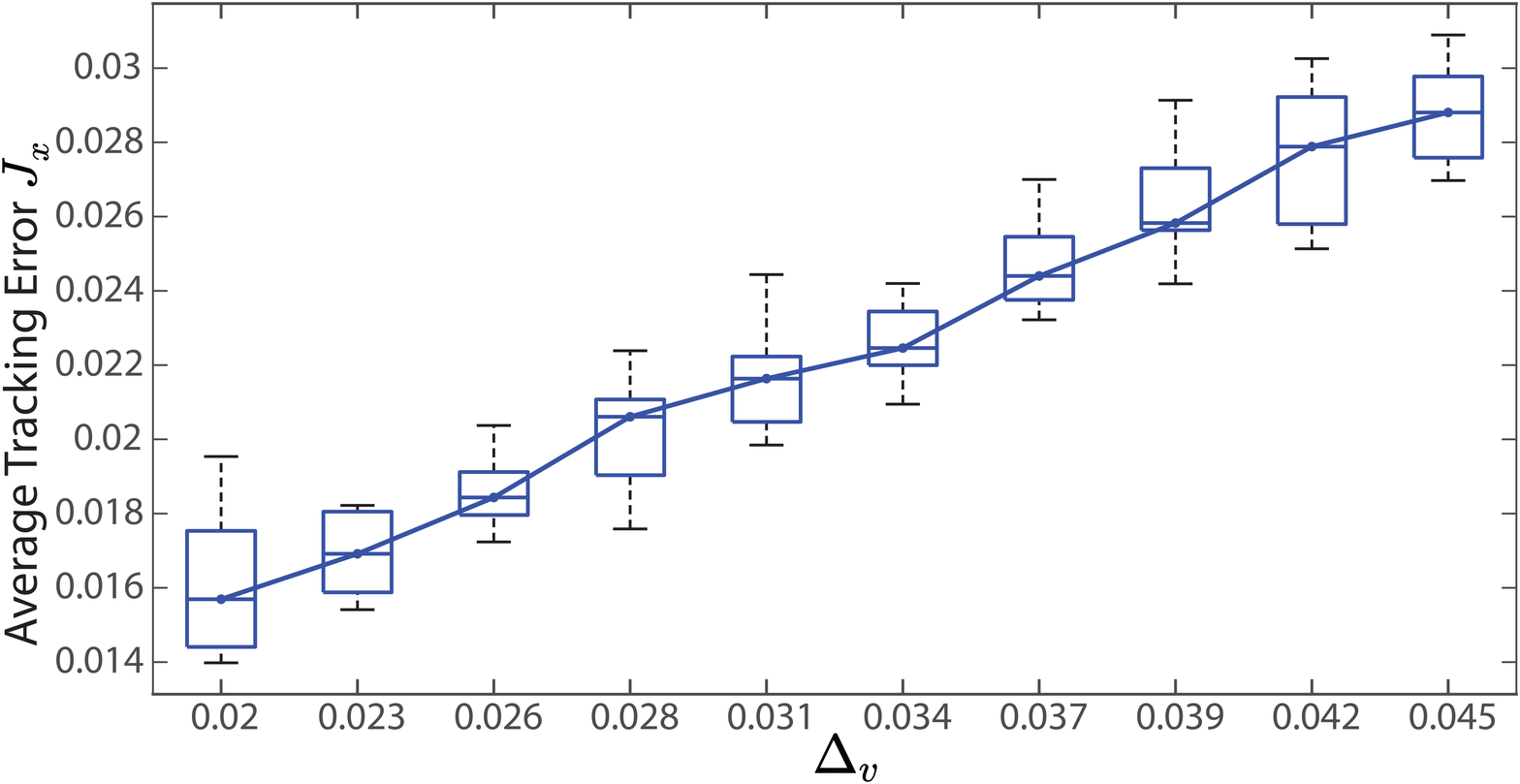}
	\caption{Performance index $J_x$ for $\Delta_v \in [0.02, 0.045].$}
	\label{fig:box_plot_total_control_cost}
\end{figure}

\begin{figure}[h!]
	\centering
	\includegraphics[width=0.75\columnwidth]{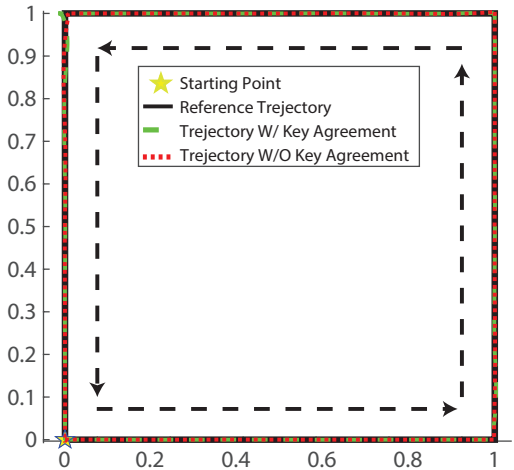}
	\caption{Square-shaped reference trajectory and robot's trajectory with (for $\Delta_v = 0.035$) and without the proposed key-agreement protocol.}
	\label{fig:robot_trajectory_experiment}
\end{figure}

\begin{figure}[h!]
	\centering
	\includegraphics[width=0.95\columnwidth]{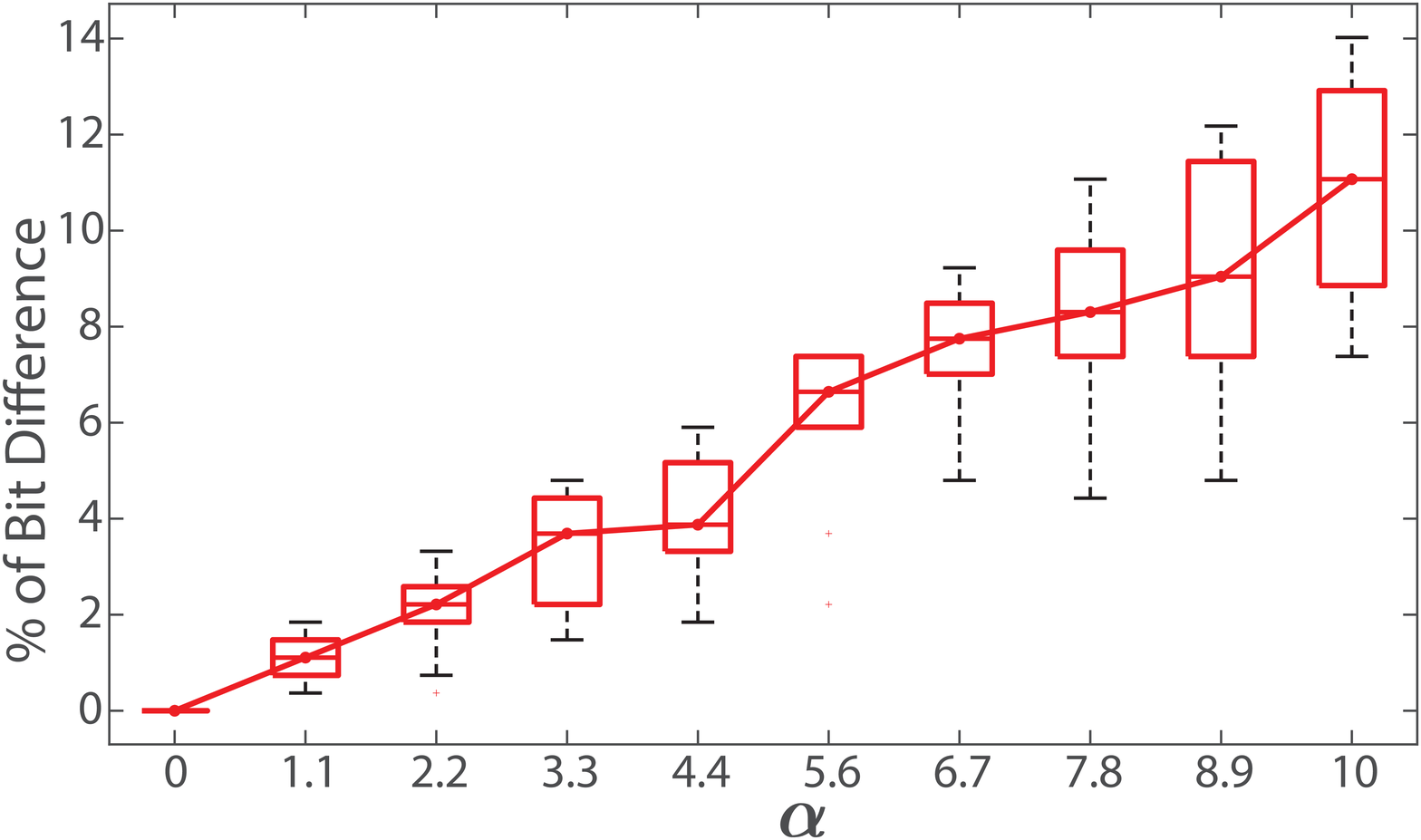}
	\caption{Bits difference (\% disagreement) between $\mathcal{K}_c$ and $\mathcal{K}_a$ for $\Delta_v=0.035$ and $\alpha \in \pm[0, 10]\%$.}
	\label{fig:Eve_recovery_error}
\end{figure}

The proposed key-agreement protocol (Algorithm~\ref{alg:proposed_protocol}) has been evaluated for 10 equally spaced value of $\Delta_v \in [0.02, 0.45].$ For each $\Delta_v,$ the experiment has been repeated 10 times and with different randomly generated keys $\mathcal{K}$ of length $345$ bits.
The obtained results are shown in Figs.~\ref{fig:Blocks_accepted_correct}-\ref{fig:Eve_recovery_error} where the shown boxplots describe the median, minimum and maximum values of each point. 

Fig.~\ref{fig:Blocks_accepted_correct} shows the percentages of accepted codewords and correctly decoded/agreed bits. The number of accepted codewords (red boxplot) increases with $\Delta_v,$ which implies that the capacity of the key agreement protocol improves with the magnitude of the state shift $\Delta_v$. Moreover, for $\Delta_v \ge 0.035$ all the accepted bits (blue boxplot) are also correct. The latter is justified by the fact that by increasing $\Delta_v,$ the distance between $u_0^D$ and $u_1^D$ increases until a point where estimation errors provoked by the process and measurement noises becomes negligible. 
%
%

On the other hand, Fig.~\ref{fig:box_plot_total_control_cost} shows the average tracking error 
$$
J_x = \frac{1}{N_s}\sum^{N_s}_{k=1}{\|[p^x(k), p^y(k)]^T - [p^x_r(k), p^y_r(k)]^T\|_2}
$$
of the robot for different values of $\Delta_v,$ where $N_s$ is the number of discrete-time steps. As expected, 
also the tracking error of the robot increase with $\Delta_v.$  Consequently, the latter suggests that the smallest value of $\Delta_v$ ensuring zero decoding errors (i.e., $\Delta_v=0.035$) should be used for key-agreement. The square-shaped reference trajectory (one lap) and the robot trajectories (one lap, single experiment) in the presence (for $\Delta_v=0.035$) and in the absence (for $\Delta_v=0$) of the proposed key-agreement protocol are shown in Fig.~\ref{fig:robot_trajectory_experiment}. There, it is possible to appreciate how the proposed key-agreement does not have a significant impact on robot reference tracking capabilities. A demo pertaining to Fig.~\ref{fig:robot_trajectory_experiment} is available at the following weblink \url{https://youtu.be/9FJkQhj8sdY}.


Finally, to test the capability of the adversary to intercept and decode the transmitted key, we have emulated an eavesdropper which has a non-perfect model knowledge $\mathcal{M}_a.$ In particular, we have assumed that the attacker knows $r,d,\mathcal{W},\mathcal{V}$ with a percentage error not superior to $\alpha.$ Moreover, for $\Delta_v=0.035,$  $10$ equally spaced values of $\alpha \in \pm[0, 10]\%$ have been considered, and for each value of $\alpha,$ $10$ experiments have been conducted with $\alpha$ randomly selected in the interval $[-\alpha,\,\alpha].$ Fig.~\ref{fig:Eve_recovery_error} reports the results of such an experiment where the y-axis shows the \% bit difference between the keys estimated by the controller ($\mathcal{K}_c$) and the adversary $\mathcal{K}_a$. As expected, the adversary conceptual channel becomes worse as the model uncertainty, i.e. $\alpha,$ increases. 



\section{Conclusion}\label{sec:conclusion}
In this paper, we have developed a key-agreement protocol for remotely controlled mobile robots without resorting to traditional cryptographic approaches.  In particular, we have leveraged the asymmetries between the controller's and adversary knowledge about the robot model to develop a key-exchange protocol based on a non-linear unknown input observer and an error correcting code mechanism.  The proposed solution has been experimentally validated using a Khepera IV differential-drive robot, and the obtained results confirmed the effectiveness of the proposed design. Future studies will be devoted develop alternative protocol capable of increasing the throughput of the key-exchange mechanism.
\bibliography{ifacconf}             
\end{document}